# A COMMENT ON THE SCALE LENGTH VALIDITY OF THE POSITION DEPENDENT DIFFUSION COEFFICIENT REPRESENTATION OF STRUCTURAL HETEROGENEITY

Molly Wolfson, Christopher Liepold, Binhua Lin* and Stuart A. Rice

The James Franck Institute, The University of Chicago, Chicago, IL 60637

*Also associated with the Center for Advanced Radiation Sources, The University of Chicago, Chicago, IL 60637

## Abstract

Experimental studies of the variation of the mean square displacement (MSD) of a particle in a confined colloid suspension that exhibits density variations on the scale length of the particle diameter are not in agreement with the prediction that the spatial variation in MSD should mimic the spatial variation in density. The predicted behavior is derived from the expectation that the MSD of a particle depends on the system density and the assumption that the force acting on a particle is a point function of position. The experimental data come from studies of the MSDs of particles in narrow ribbon channels and between narrowly spaced parallel plates, and from new data, reported herein, of the radial and azimuthal MSDs of a colloid particle in a dense colloid suspension confined to a small circular cavity. In each of these geometries a dense colloid suspension exhibits pronounced density oscillations with spacing of a particle diameter. We remove the discrepancy between prediction and experiment using the Fisher-Methfessel interpretation of how local equilibrium in an inhomogeneous system is maintained to argue that the force acting on a particle is delocalized over a volume with radius equal to a particle diameter. Our interpretation has relevance to the relationship between the scale of inhomogeneity and the utility of translation of the particle MSD into a position dependent diffusion coefficient, and to the use of a spatially dependent diffusion coefficient to describe mass transport in a heterogeneous system.

## Introduction

It has become common practice to describe diffusive motion in a system that has structural heterogeneity with a position dependent diffusion coefficient [1-4]. The position dependent local diffusion coefficient is, typically, obtained from measurement of the mean square displacement (MSD) of a probe particle and translated into a *d*-dimensional local diffusion coefficient using the Einstein relation $\langle(\Delta r)^2\rangle$ =*2dDt*. That defined diffusion coefficient is then used in the canonical diffusion equation to describe mass flow in the system. This short paper is concerned with understanding a discrepancy: Experimental studies [5,6] of the position dependence of the MSD of a particle in a confined quasi-two-dimensional (q2D) colloid suspension that exhibits density variations on the scale length of the particle diameter are not in agreement with the prediction that the spatial variation in MSD should mimic the spatial variation in density. We are interested in resolving that discrepancy, i.e., in determining when and on what scale length a position dependent diffusion coefficient is an appropriate representation of the interaction of a probe with its surroundings.

Brownian motion is an apt descriptor of the particle dynamics in soft matter systems, both physical and biological, because of the important influence on the particle motions of thermal fluctuations. In a dilute colloid suspension that is homogeneous and unbounded, on a time scale that is large compared with the inverse of the frequency of particle displacement, the distribution of single particle displacements is Gaussian and the MSD of a particle is linear in time and independent of location in the system. In contrast, when a colloid system is both dense and inhomogeneous on some scale the single particle MSD has, typically, different time dependences as *t* increases, the distribution of displacements is, typically, non-Gaussian, and the relationship between the MSD of a particle and the macroscopic diffusion coefficient is more complex. For such systems a diffusion coefficient can be defined by $D(\mathbf{r},t) = \frac{1}{2d}\frac{\partial\langle(\Delta r)^2(t)\rangle}{\partial t}$. So-called anomalous diffusion is characterized by a MSD that has the form $\langle(\Delta r)^2(t)\rangle \sim D(\mathbf{r})t^\alpha$. Both the position dependent $D(r)$ and the time scaling exponent $\alpha$ that defines the regions of sub-diffusive ($\alpha<1$), super-diffusive ($\alpha>1$) and normal diffusive ($\alpha=1$) behavior, are monitors of the structural inhomogeneity of the system. Note that this definition of the diffusion coefficient presumes the existence of, but does not identify, the character of the structural inhomogeneity in the crowded system. We will comment further on the character of the time dependence of the single particle MSD and its relationship with $D(\mathbf{r},t)$ along long trajectories in a dense q2D colloid suspension in the next Section.

In a number of cases it is found that although the particle MSD in a system grows linearly with time, the distribution of particle displacements is not Gaussian [7,8]. The characteristic features of diffusion in such cases are captured by the so-called diffusing diffusivity model [9], which posits the existence of a distribution of regions in the system in which conventional diffusion with a Gaussian distribution of displacements occurs, but with the diffusivity different in each region (hence position dependent) subject to the exponential distribution $P_D(D) = \langle D \rangle^{-1} \exp\left(-\frac{D}{\langle D \rangle}\right)$ with mean diffusion coefficient <D>. An example of this type is provided by diffusion of a particle near a wall in a dilute colloid suspension. In this case the diffusion coefficient has a known spatial dependence (distance from the wall) that is generated by hydrodynamic interaction between the particle and the wall, and an elegant recent study [10] has verified the deviation of the particle displacement distribution from Gaussian that is predicted by the diffusing diffusivity mechanism. This is a case in which the extra hydrodynamic force on the particle that is generated by the boundary condition at the wall is properly represented as a point function of position.

Another category of systems for which introduction of a space dependent diffusion coefficient is appropriate consists of those with boundary conditions that vary the confinement of the system with position, e.g. quasi-one-dimensional narrow channels with rippled walls. In such systems a position dependent, time independent, entropic force is exerted on a diffusing particle, on which we will comment further in the next Section.

We argue that whether or not a position dependent diffusion coefficient is an apt representation of the influence on the single particle motion of inhomogeneity in a system depends on the nature of confinement of the system, on the character and scale of the inhomogeneity, and on the nature of the force exerted on a diffusing particle. We ask three related questions:

1. For what length scale of a spatial inhomogeneity in a system is it appropriate to introduce a space dependent diffusion coefficient?
2. For what category of forces acting on a particle is it appropriate to introduce a space dependent diffusion coefficient?
3. If $D(\mathbf{r})$ is known, can it be inverted to yield information about the inhomogeneity in the system?

**The Discrepancy**

Because of its simplicity, we consider first a quasi-one-dimensional system that is confined within boundaries that have structure that has scale length that is large compared with a particle diameter, e.g. a ribbon with periodically wavy walls. In this case it is the variation in boundary shape that generates a position dependent entropic force that acts on a diffusing particle. The asymptotic long-time diffusion in a channel with variable cross section $w(x)$ is found to be properly described by use of an effective position dependent diffusion coefficient and an evolution equation of the form

$$\frac{\partial \rho(x,t)}{\partial t} = \frac{\partial}{\partial x}\left(w(x)D(x)\frac{\partial}{\partial x}\left[\frac{\rho(x,t)}{w(x)}\right]\right) \qquad (1)$$

Zwanzig [11], Reguera and Rubi [12], and Kalinay and Percus [13] have studied how $D(x)$ is related to the cross section of the channel. The latter find, to lowest order in $w(x)$ and its derivatives, that

$$D(x) = D_0 \arctan\left(\frac{1}{2}\frac{dw(x)}{dx}\right)\left(\frac{1}{2}\frac{dw(x)}{dx}\right)^{-1} \qquad (2)$$

The effective one-dimensional diffusion equation can be shown to be equivalent to a description of particle motion with a one-dimensional Langevin equation with one particle potential

$$V(x) = -k_B T \ln(w(x)) \qquad (3)$$

The forms of $D(x)$ and $V(x)$ displayed are valid when $w(x)$ varies slowly on the scale length of the particle diameter. Note that $V(x)$ is a point function of position and that, because the boundary is fixed, $V(x)$ and $D(x)$ are independent of time.

In the case just considered the origin of the force acting on the particle is the variation of the width along the quasi-one-dimensional channel. In a sense, that force can be considered to be created by an external source. A different situation is created when the equilibrium density variation of the system along some coordinate is obtained, internal to the system, from an ensemble average. That density variation then generates a stationary potential of mean force that acts on a diffusing particle. An example is provided by the MSD of a particle in the inhomogeneous liquid confined between two flat plates. Mittal, Truskett, Errington and Hummer [14] analyzed such a system, specifically the diffusion of hard spheres confined between closely spaced parallel hard plates. They start with the assumption that the probability density for finding a particle at position $z$, $\rho(z,t)$, satisfies the Smoluchowski equation

$$\frac{\partial \rho(z,t)}{\partial t} = \frac{\partial}{\partial z}\left\{D_z(z)e^{-F(z)/k_B T}\frac{\partial}{\partial z}\left[e^{F(z)/k_B T}\rho(z,t)\right]\right\} \qquad (4)$$

with position dependent diffusion coefficient $D_z(z)$. The free energy profile along $z$ is assumed to have the form $F(z) = -k_B T \ln \rho(z)$ with $\rho(z)$ the equilibrium density profile. Note that the force acting on the particle, derived from $F(z)$ via the spatial variation of $\rho(z)$, is a point function of $z$. Mittal et al predict that at high packing fraction, for which stratification of the density along the $z$ direction is prominent, the variation of $D_z(z)$ mimics the structure of $\rho(z)$. A related analysis, by Colmenares, Lopez and Olivares-Rivas [15], studied diffusion of Lennard-Jones particles confined between smoothed planar Lennard-Jones walls. They start by converting the Langevin equation of motion into an equation for the MSD of a particle with the mean value computed for particles constrained to be in a single layer of width $L$ in the $z$ direction; the force on a particle within that layer is taken to be a constant and the noise within that layer is taken to have zero average value. With some further approximations they predict that the variation of $D_z(z)$ mimics the structure of $\rho(z)$. This analysis also assumes that the force on a particle that appears in the Langevin equation is a point function of position.

However, the available experimental data for systems with equilibrium stratified density distributions do not agree with the predictions obtained from the analyses of Mittal et al and Colmenares et al. Experimental studies by Wonder, Lin and Rice [5], of the single particle MSD in an inhomogeneous monolayer colloid suspension confined to a ribbon channel, do not show any correlation between the MSD of a particle and the well-defined peaks in the transverse density distribution (See Figs. 2 and 3 of Ref. 5). Similarly, experimental studies by Edmond, Nugent and Weeks [6] of the single particle MSD in colloid suspensions confined between two flat plates with separations of a few particle diameters do not show any correlation between the MSD of a particle and the well-defined peaks in the transverse density distribution. We add to this data set with a report, in the appendix to this paper, of the results of a study of the azimuthal and radial single particle MSDs of a particle in a dense q2D colloid suspension confined in a small circular cavity. In this suspension the colloid density exhibits large amplitude oscillations as a function of radial distance from the cavity wall. It is found that the particle MSDs do not mimic the local structure in the liquid. The azimuthal MSD is sensibly independent of the variation in particle density along the radius of the cavity and the radial MSD is only weakly dependent, close to the cavity boundary, on the radial variation in particle density.

To help understand the discrepancy between theory and experiment just described for systems with inhomogeneous density distributions with scale length the same as the particle diameter it is useful to consider the relationships between structural heterogeneity and the particle MSD over short and long trajectories. The vehicle for our analysis is the particle MSD in a dense unbounded q2D colloid suspension [16]. In this system the overall distribution of displacements deviates from Gaussian form, and in different intervals the MSD has different time dependences. Specifically, it is found that there is an intermediate time domain in which motion is sub-diffusive, $\alpha < 1$, bracketed at shorter and longer times by domains in which motion is diffusive, $\alpha = 1$, with different slopes. The connection between the time dependences of the MSD in these time domains and system heterogeneity is established by examination of the images of particle configurations in the colloid suspension. These images reveal that the temporal behavior of $\langle r^2(t) \rangle$ is associated with the existence of a patchwork of transient structural ordering in the system (see Fig. 5 of [16]). In an unbounded q2D dense colloid assembly at equilibrium, fluctuations in the high-density liquid generate spatial configurations that consist of small transiently ordered domains separated by narrow disordered boundaries. The transient ordering has a finite lifetime because of exchanges of particles between the ordered and disordered patches, but successive images show the same overall fractions of transiently ordered domains and disordered boundaries. The short time motion of a particle in a transiently ordered domain is constrained and over-damped but fully two-dimensional ($\alpha = 1$). The motion of a particle in a disordered boundary between ordered domains has considerable one-dimensional file-server character ($\frac{1}{2} < \alpha < 1$) (See Fig. 10 of [16]). At long time the MSD is again linear in time. The MSD of a particle can be characterized with three simultaneous competing relaxation processes each of which generates a Gaussian distribution of displacements. For an interval that is shorter than the time required by an over-damped particle to move a significant fraction of a particle diameter the particle displacements occur within an ordered domain inside a cage of fluctuating neighbors. At somewhat longer time the file-server-like contribution arises from correlated motion in the disordered sensibly linear boundary regions. At very long time there are contributions to $\langle r^2(t) \rangle$ from infrequent large displacements, of the order of a particle diameter in length (See Fig. 16 of [16]). These large displacements are not ballistic; they are associated with density fluctuations that reduce the coordination number of the surroundings of a particle. The overall picture that emerges supports the view that the single particle MSD can depend on location in a particular subset of time-adjacent configurations but, because the exchanges of particles between the ordered and disordered domains generates an average over the configurations, the MSD is not a stationary function of position in the liquid and cannot be meaningfully converted to a position dependent diffusion coefficient.

**Commentary**

What is the source of the discrepancy between the observed behavior of the diffusion coefficient in a medium with equilibrium inhomogeneity on the scale length of the particle diameter, and that predicted via analyses using the Smoluchowski or Langevin equations? And, why is the MSD of a particle in a system with inhomogeneity on the scale length of the particle diameter independent of the density distribution?

In a bulk suspension the diffusion coefficient is a strong function of the colloid density. The conventional local density representation of the equilibrium properties of an inhomogeneous system posits that the density is a point function of the system that satisfies the equation of state. Then, in a stratified suspension, one is led to the expectation of a corresponding variation with stratum density of the diffusion coefficient. We interpret our observation vis a vis the insensitivity of the MSD to position in a region with strong density variation on the scale length of a particle diameter using the Fischer-Methfessel representation of the local density in an inhomogeneous fluid [17]. Fischer and Methfessel pointed out that to sustain a density gradient in an inhomogeneous liquid there must be a balancing force that is not captured by representing the local density as a point function. They showed that to lowest order the source of the force is, typically, interaction with nearest neighbors of a molecule, and therefore they defined the local density as an average over a volume with radius one particle diameter. It has been shown that this approximation provides a good description of the pair correlation function in the strongly inhomogeneous transition region in the liquid-vapor interface [18]. The application to diffusion follows from the observation that the friction coefficient for a particle is determined by the force-force correlation function. Both the direct force acting on a particle and the force generated by the hydrodynamic interactions between colloid particles are determined by the pair correlation function. Applying the Fischer-Methfessel approximation to the description of a system with stratification with scale length of a particle diameter system requires defining the local density in the volume determined by averaging over neighboring strata. Noting that the minima and maxima of the strata densities are approximately equal and spaced by approximately one particle diameter, this averaging effectively removes the density dependence of the pair correlation function and, finally, the density variation of the friction coefficient.

Thus, our answer to question (1) posed in the Introduction is that the use of a position dependent diffusion coefficient to reflect the influence of system inhomogeneity on macroscopic transport of mass is valid only when the inhomogeneity is both stationary in time and of scale length large compared with the particle diameter in order that the force acting on the diffusing particle can be represented as a point function of position. When the inhomogeneity arises from equilibrium local structure on the scale length of the particle diameter, the force acting on a particle is delocalized over several particle diameters, the MSD of a particle then does not mirror the inhomogeneity and the diffusion coefficient obtained from the MSD is not position dependent.

Our answer to question (2) has been stated several times: We argue that a one-to-one correspondence between a position dependent diffusion coefficient and a spatially distributed inhomogeneity requires that the force acting on a particle is a point function of position. This condition is met for an inhomogeneity with spatial extent large compared to a particle diameter but not when the spatial variation of the source of the force has the same scale as the particle diameter.

Assuming that $D(\mathbf{r})$ is known, an answer to question (3) depends on having an analytic representation of the relation between the force acting on a particle and the spatial distribution of the source of that force, i.e. the analog of the relationship between wall shape $w(x)$ and $D(x)$ in the quasi-one-dimensional system. However, when the structure in the medium is not derived from the shape of the walls we do not know the functional form of that relation, and the inversion of $D(\mathbf{r})$ to determine the functional form of the underlying structural inhomogeneity is not possible.

## Acknowledgements

We thank Prof. Suriyanarayanan Vaikuntanathan and Prof. Haim Diamant for fruitful conversations. This work was supported by the NSF-MRSEC (DMR-1420709) Laboratory at the University of Chicago and by a grant from the Camille and Henry Dreyfus Foundation (SI-14-014). B.L. acknowledges support from ChemMatCARS (NSF/CHE-1346572).

## Appendix

The trajectories of colloid particles confined to one layer in circular cavities with diameters 40 and 75 $\mu m$ were studied using digital microscopic techniques and equipment described in detail elsewhere [19]. The cavities were prepared by pouring uncured polydimethylsiloxane (PDMS, Sylgard 184) on a silica wafer that was lithographically etched with the negatives of the desired cavities shapes. The cured PDMS wafer was then stripped from the silica mold. All of the cavities studied have a depth of 3 $\mu m$; they were filled by placing approximately 10 $\mu m$ of an aqueous suspension of 1.57(2) $\mu m$ silica spheres (Duke Standards 8150) on top of a region of the PDMS wafer, followed by coverage with a thin glass cover slip. The cover slip prevented the confined aqueous colloid suspension from evaporating for a time adequate to conduct the trajectory measurements and also inhibited the generation of flows during the course of the experiment. The filling of the cavities was accomplished by the gravitational settling of the colloid particles (mass density = 2.2g/cm$^3$), a process that is not directly controlled in the preparation procedure. For that reason the colloid packing density in a cavity was not predetermined, and the reported values of that density were determined from the digital images during the data analysis. Different colloid packing densities were generated by dilution of the original vendor suspension with deionized water. The samples studied were selected to be free of particles stuck to the confining windows and walls.

The measured trajectories were obtained from digital video images collected with an Olympus BX51 System Microscope and a Leica DFC310 FX digital color camera. The videos of the suspensions were taken with a frame rate of 17 frames per second for 6-12 minutes. The displacement of a particle was decomposed into the motion along a radius of the cavity, determined by the center of the cavity and the initial particle position, and the motion perpendicular to the radial direction (the azimuthal direction). The measured MSD is linear in time for about 10 – 15 times the time required for a particle to diffuse a distance on one particle diameter after which the time dependence starts to be sub-linear (Figs. 1 and 2); we are concerned only with the linear time regime. When desired, specific radial annuli of the cavities were labeled digitally. With this labeling it can be determined that a particle initially in an annulus remains in that annulus for the duration of the measurement, thereby permitting determination of the azimuthal MSDs for such constrained particles. Fig. 3 displays a frame from a video recording and a plot of the trajectories of the particles over an interval of 1000 frames. The image and the track pattern show the influence of structured colloid packing close to the boundary of the cavity.

A summary of our experimental findings is presented Figs. 4 and 5: Fig. 4a displays the radial and azimuthal MSDs/s as a function of radial position in a q2D colloid suspension with packing fractions 0.72 in a cavity with diameter of 40 $\mu m$, Fig 4b displays the radial colloid density distribution in that cavity, and Figs. 5a and 5b show the corresponding data for a q2D colloid suspension with packing fraction 0.62 in a cavity with diameter of 75 $\mu m$. The most important inference obtained from these data is the lack of any structured radial variation in the measured MSDs that mimics the radial density distributions. The azimuthal MSDs are sensibly independent of radial position and the radial MSDs show only a weak unstructured decrease in magnitude on progressing from the center of the experimental chamber to the confining wall.

## Figures

Figure 1

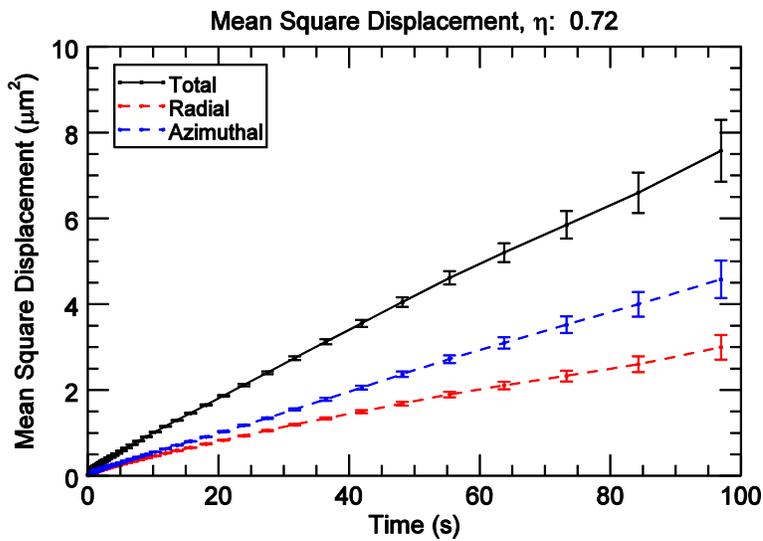

Caption: (color online) the MSD of a colloid particle in the confined q2D suspension with $\eta = 0.72$ and a cavity diameter of about 40 μm.

Figure 2

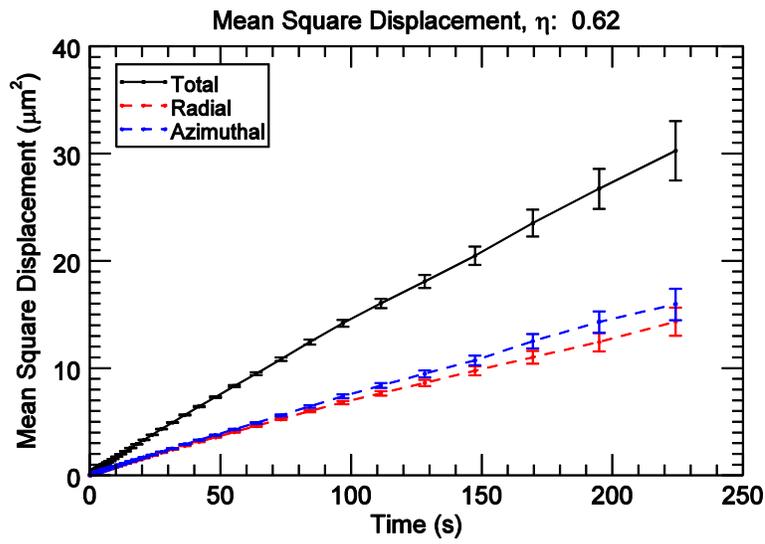

Caption: (color online) the MSD of a colloid particle in the confined q2D suspension with η = 0.62 and a cavity diameter of about 75 μm.

Figure 3

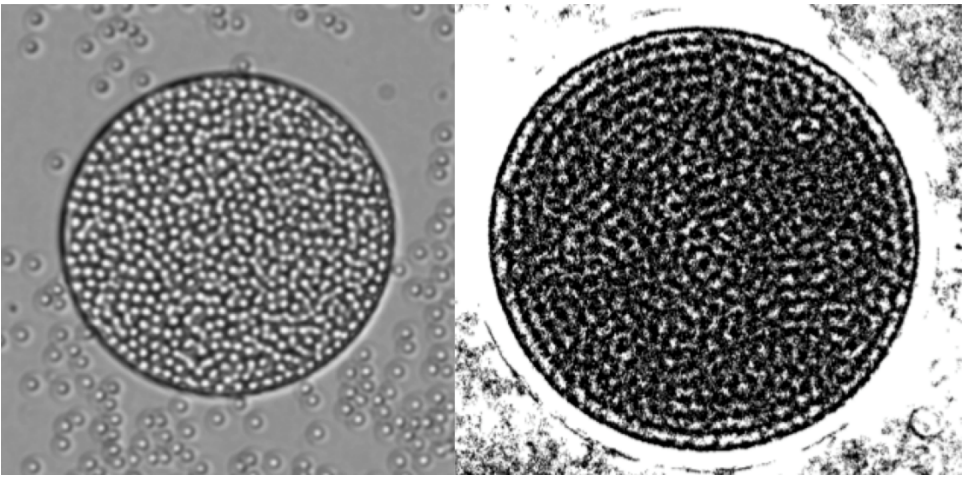

Caption: A frame from the experimental video data (left) and a plot of the tracks of the particle trajectories over 1000 consecutive frames with η = 0.72.

Figure 4 (a)

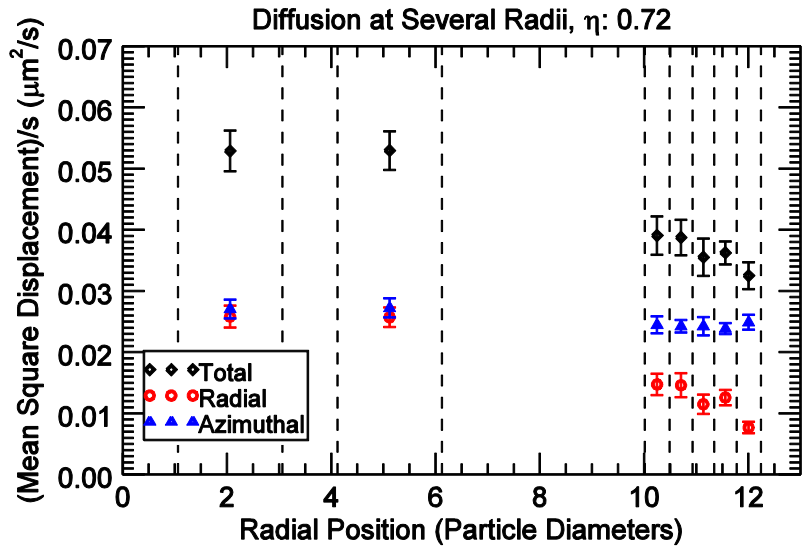

Figure 4 (b)

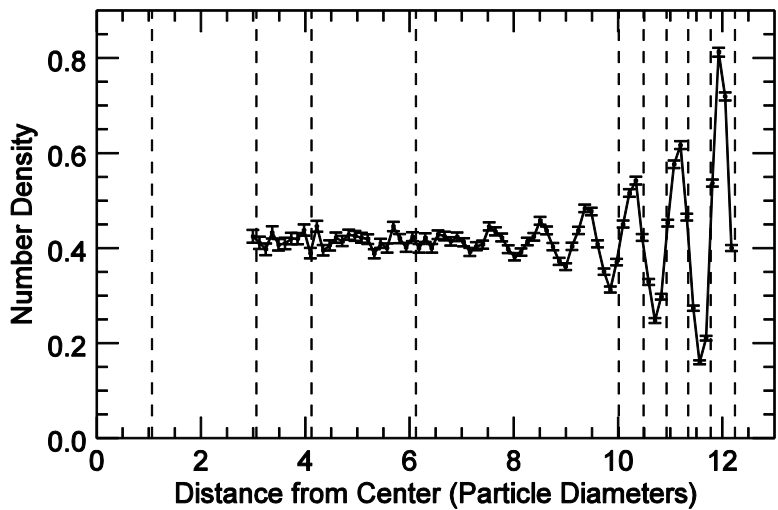

Caption: (a) (color online) the MSD of a colloid particle as a function of radial position and (b) the colloid density as a function of radial position in the confined q2D suspension with $\eta=0.72$ and a cavity diameter of 40 μm.

Figure 5 (a)

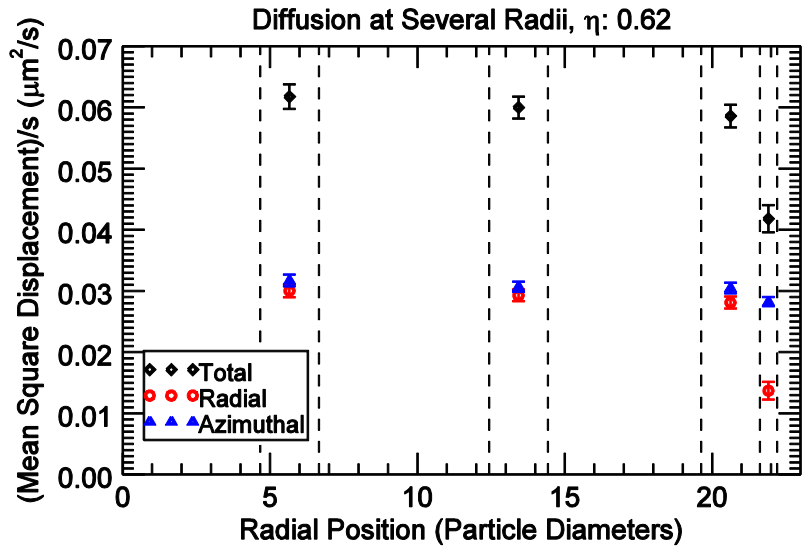

Figure 5 (b)

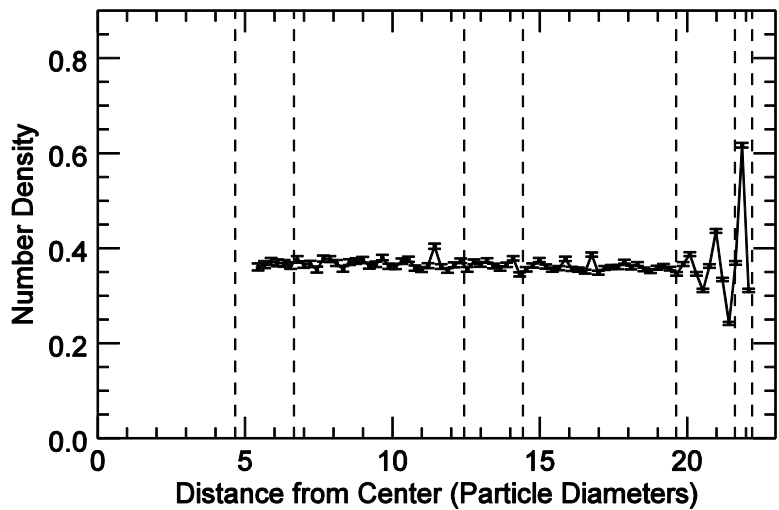

Caption: (a) (color online) the MSD of a colloid particle as a function of radial position and (b) the colloid density as a function of radial position in the confined q2D suspension with η=0.62 and a cavity diameter of 75 μm.